\documentclass[aps, prl, amsmath, amsfonts, twocolumn, superscriptaddress, 10pt]{revtex4-2}
\usepackage[caption=false]{subfig}  
\usepackage{graphicx}
\usepackage{upgreek}

\DeclareMathOperator*{\argmin}{arg\,min}
\newcommand{\norm}[1]{\left\lVert#1\right\rVert}

\newcommand{\g}{\boldsymbol{g}}
\newcommand{\ghat}{\boldsymbol{\widehat{g}}}
\newcommand{\ghati}{\widehat{g}}
\newcommand{\x}{\boldsymbol{x}}
\newcommand{\X}{\mathbf{X}}
\newcommand{\cvec}{\boldsymbol{c}}
\newcommand{\Svec}{\mathbf{\Sigma}}

\begin{document}

\title{Local Clustering and Global Spreading of Receptors\\ for Optimal Spatial Gradient Sensing}
\author{Albert Alonso}
\affiliation{Niels Bohr Institute, University of Copenhagen, Denmark}
\author{Robert G. Endres}
\affiliation{Department of Life Sciences and Centre for Integrative Systems Biology and Bioinformatics, Imperial College, London, United Kingdom}
\author{Julius B. Kirkegaard }
\email[Correspondence email address: ]{juki@di.ku.dk}
\affiliation{Niels Bohr Institute, University of Copenhagen, Denmark}
\affiliation{Department of Computer Science, University of Copenhagen, Denmark}
\date{\today}

\begin{abstract}
Spatial information from cell-surface receptors is crucial for processes that require signal processing and sensing of the environment.
Here, we investigate the optimal placement of such receptors through a theoretical model that minimizes uncertainty in gradient estimation.
Without requiring \textit{a priori} knowledge of the physical limits of sensing or biochemical processes, we reproduce the emergence of clusters that closely resemble those observed in real cells.
On perfect spherical surfaces, optimally placed receptors spread uniformly.
When perturbations break their symmetry, receptors cluster in regions of high curvature, massively reducing estimation uncertainty.
This agrees in many scenarios with mechanistic models that minimize elastic preference discrepancies between receptors and cell membranes.
We further extend our model to motile receptors responding to cell-shape changes and external fluid flow, demonstrating biological relevance of our model.
Our findings provide a simple and utilitarian explanation for receptor clustering at high-curvature regions when high sensing accuracy is paramount.
\end{abstract}
 
\maketitle

Cells rely on their ability to detect and respond to environmental cues for essential biological processes including chemical gradient sensing in chemotaxis, wound healing, and embryonic development.
It is only logical that the spatial information gathered by the cells is highly dependent on the positioning of receptors on the cell surface, and consequently on cell shape itself.
Hence, cells can actively influence sensing performance, as exemplified by yeast cells localizing receptors such as Ste2 to shmoo tips when projecting up mating pheromone gradients~\cite{moore_yeast_2013}.
In many biological systems, receptors are observed to cluster, particularly in regions of high membrane curvature, avoiding the mismatch of the preferred curvatures of membrane and embedded proteins~\cite{endres_polar_2009}, or due to biochemical constraints such as attractive protein-protein interactions, facilitating chemical reactions and efficient signaling~\cite{lai_cooperative_2005,  mcandrew_tem_2004, ames_collaborative_2002}.
Receptor clustering also occurs purely by stochastic self-organization, such as driven by cell growth, receptor diffusion, and capture~\cite{thiem_stochastic_2008}.
These studies explain the mechanistic algorithms by which clusters are formed.
However, what other evolutionary advantages do receptor localization and clustering provide to cells with, especially when confronted with noisy spatial signals?

Here, we develop a simple theoretical model to study the relationship between receptor distribution, cell shape, and gradient sensing efficiency in arbitrary 3D geometries.
By inducing shape perturbations, we observe that clusters emerge naturally to minimize the uncertainty of the spatial gradient estimation, without any need to explicitly invoke receptor interactions or membrane information.
The clusters that result from our model are localized in high curvature regions in agreement with both experimental studies on prokaryotic~\cite{jones_positioning_2015, koler_long-term_2018, kentner_determinants_2006} and eukaryotic~\cite{cai_t_2022, wan_origins_2021, koldso_lipid_2014, yang_membrane_2022} systems, as well as theoretical studies that model receptor interactions and membrane curvature sensing~\cite{lin_curvature-induced_2024, wang_self-organized_2008, draper_origins_2017}.
Finally, we extend the model in two key directions:
First, we study how receptors move in response to dynamic changes in the cell body, such as cellular protrusion, e.g. actin-driven  pseudopods~\cite{sanchez_ligand-independent_2023, alonso_persistent_2024}.
Second, we elucidate how receptor localization ideally responds to flow of the surrounding fluid, which produces intricate patterns that may have implications in the context of synthetic biology~\cite{bi_engineering_2016}, biosensor designs~\cite{chircov_biosensors--chip_2020}, and robotics at the physical limits of sensing~\cite{deng_amoeboid_2023, zhou_bioinspired_2024}.

\vspace{2em}
Consider a cell placed inside a three-dimensional diffusive environment with a linear gradient concentration profile
\begin{equation}\label{eq:concentration-profile}
   c(x, y, z) = c_0 + g_x x + g_y y + g_z z = \x \cdot \g
\end{equation}
where $\x = \begin{pmatrix} 1, x, y, z \end{pmatrix}$ is extended positional coordinates and $\g = \begin{pmatrix} c_0, g_x, g_y, g_z \end{pmatrix}$ is the background concentration and gradient values.

We assume receptors to be \textit{perfect instruments}~\cite{berg_physics_1977} for counting molecules, and the cellular internal mechanisms of biochemical signaling pathways to perfectly process the information from the $n$ surface receptors to distill the best gradient estimator $\ghat$.
Assuming a shallow gradient ($R |\nabla c| \ll c_0$), we can model a cell measurement as a multivariate Gaussian,
\begin{equation}
    P(\cvec) = \frac{1}{\sqrt{(2 \pi)^n | \Svec |}} e^{- \frac{1}{2} \left(\cvec - \X \cdot \g \right)^{T} \Svec^{-1} \left(\cvec - \X \cdot \g \right)},
\end{equation}
where $\Svec$ is the covariance matrix between measurements $\cvec = \left( c_1, \dots, c_n \right)$, and $\X$ is a positional matrix in $\mathbb{R}^{n\times 4}$ where each row is the extended position vector of a receptor at the surface of the cell.
Assuming that the concentration fluctuations are diffusion-limited, we model the individual uncertainty of each measurement with Berg and Purcell noise~\cite{berg_physics_1977}, letting each receptor make an average estimate over measurement time $\tau$.
This reduces the measurement variance \cite{berg_physics_1977, endres_accuracy_2008}, but also introduces the possibility of rebinding, thus causing the measurements of nearby receptors to be correlated.
Taking all this into account we use
\begin{equation}\label{eq:covariance-matrix}
   \Sigma_{i,j} = \frac{2 a^3 V}{D\tau(5 a + 6 \Delta)} (\x_i + \x_j)  \cdot \g.
\end{equation}
where $\Delta=\norm{\x_i-\x_j}$ is the distance between receptors, $D$ is the molecular diffusion coefficient, $\tau$ the cell measuring time, $a$ and $V$ the receptor effective radius and volume, respectively (see~\cite{noauthor_supplementary_nodate} for a detailed derivation).
In our formulation, $a$ determines the typical correlation distance of the receptors.

We are interested in optimizing the uncertainty of the gradient estimation as a function of the receptor locations.
The Cramer-Rao bound provides a lower bound on the variance of any unbiased estimator~\cite{aquino_know_2016, hu_how_2011, hopkins_chemotaxis_2020},
\begin{equation}\label{eq:cramer-rao-bound}
\text{Cov}(\ghat) \geq \mathcal{I}(\X, \ghat) ^ {-1},
\end{equation}
where $\mathcal{I}$ is the Fisher Information Matrix, which for our multivariate Gaussian evaluates to
\begin{equation}\label{eq:FIM}
    \mathcal{I}_{m,k} = \frac{\partial c^T}{\partial \ghati_m} \Sigma^{-1} \frac{\partial c}{\partial \ghati_k} + \dfrac{1}{2}\text{tr}\left( \Sigma^{-1} \frac{\partial \Sigma}{\partial \ghati_m} \Sigma^{-1} \frac{\partial \Sigma}{\partial \ghati_k}\right).
\end{equation}
where we use Eq.~\eqref{eq:concentration-profile} with the real gradient $\g$ to obtain the expected concentration $c$,
and, using Eq.~\eqref{eq:covariance-matrix}, 
\begin{equation}\label{eq:partial-derivative-covariance}
   \frac{\partial \Sigma_{i,j}}{\partial \ghati_k} = \frac{2 a^3 V}{D\tau(5 a + 6 \Delta)} \left( X_{i,k} + X_{j,k} \right).
\end{equation}

Our aim is to find the optimal spatial configuration $\X^\ast$ of the $n$ surface receptors to minimize the uncertainty of the gradient estimations,
\begin{equation}\label{eq:ghat}
    \delta \ghati (\X) = \text{tr}(\text{Cov}(\ghat; \X))
\end{equation}
such that the optimal spatial distribution fulfills $\mathbin{\X^\ast=\argmin_{\mathbf{X}\in\mathcal{S}^n}\!\left(\delta \ghati \right)}$ where $\mathcal{S}$ is the surface of the cell.
We solve the optimization problem with differentiable programming and local gradient descent.

\begin{figure}[!tb]
    \centering
    \includegraphics[width=3.3in]{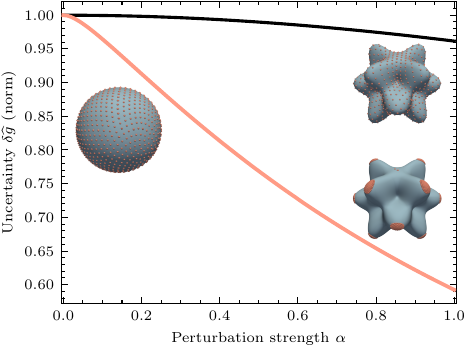}
    \caption{
    Uncertainty of gradient estimation by the cell, normalized by the initial value of $\delta \ghati(X_0)$, of a spherical cell perturbed by the harmonic $Y_6^4$ to form protrusions, in the case of uniformly spread receptors (black) or when placed at their optimal location for each $\alpha$ (orange).}
    \label{fig:clusters}
\end{figure}

\vspace{2em}
Using spherical coordinates ($r$, $\theta$, $\phi$), we define the cell surface by $r = f(\theta, \phi)$.
Our results are valid for any positive cell-shape function $f$, and we exemplify using spherical harmonics,
\begin{equation}\label{eq:radius-harmonics}
    r(\theta, \phi) = r_0 + \alpha \times \text{Re}\left[Y_{\ell}^{m}(\theta, \phi)\right],
\end{equation}
which have previously been shown to be effective in describing cell shapes \cite{cavanagh_t_2022}, in particular perturbations during cell migration (see~\cite{noauthor_supplementary_nodate} for more complex shapes).
$Y_{\ell}^{m}$ is the spherical harmonic of degree $\ell$ and order $m$, $r_0$ is the base spherical radius of the cell~\footnote{We set $r_0=1$ thus setting our units in terms of cell size.}, and $\alpha$ is a free parameter that controls the strength of the harmonic perturbations, meant to induce surface irregularities.

Using the Golden Spiral algorithm (``the Fibonacci sphere'')~\cite{hardin_comparison_2016}, we initialize the receptors to be approximately uniformly distributed on the surface before optimization.
In the case of the unperturbed cell ($\alpha=0$), this provides a very evenly distributed distribution along the surface (Fig.~\ref{fig:clusters}).
We note that the more perturbed the cell surface is from a sphere, the less perfectly distributed the receptors will be at the start of the optimization.

As the cell has to estimate the gradient with three independent directional components, the optimal distribution requires positioning the receptors to enable gradient estimation in any direction.
On a sphere this results in uniformly spaced receptors.
However, we observe that $\delta \ghati$ is very insensitive to the precise localization of the receptors on a sphere, as long they are mostly spread.
For symmetric cell shapes, we thus expect even small effects coming from biochemical and physical factors would dominate the minor changes to receptor efficiency.

Once the sphere is perturbed and the symmetry is broken ($\alpha>0$), the optimal distribution is no longer uniform.
Instead a clustered distribution emerges, where receptors move to gather at the tips of protrusions (Fig.~\ref{fig:clusters}).
This clustering aligns with experimental observations~\cite{stamou_membrane_2015, rosholm_membrane_2017}, suggesting that receptors aggregate in areas of higher curvature.
Notably, this emerges simply by breaking the surface symmetry.
Intuitively, receptors must be placed to maximize the spatial spread in all directions, which they achieve by collapsing to the tips of the protrusions.
Spatial correlations in receptor binding however penalizes the proximity between the receptors and thus results in clustering of finite size~(see~\cite{noauthor_supplementary_nodate} for a simple one-dimensional demonstration).

By modifying the value of $\alpha$, we can smoothly perturb the spherical shape of the cell.
Previous studies have considered the impact of cell shape and size on gradient estimation~\cite{mou_optimal_2024, nakamura_gradient_2024, baba_directional_2012, tweedy_distinct_2013, alonso_learning_2024}.
These studies find that elongated or protruded shapes can improve gradient estimation without accounting for the repositioning of receptors.
This type of effect is reproduced in Fig.~\ref{fig:clusters} (black curve), showing an improvement in estimation accuracy.
Additionally placing the receptors in their optimal locations leads to a huge additional improvement (orange curve), thus showing the true potential of shape deformations on gradient estimation.
In other words, any estimation of the improvement in accuracy resulting from deforming cell shapes will drastically underestimate the effect if it does not account for receptor localization.

\vspace{2em}
Previous work has shown that clustering occurs predominantly in high curvature regions~\cite{endres_polar_2009, duke_equilibrium_2009}, in particular in regions of high negative mean curvature $H$, corresponding to receptors following gradient $\nabla H$.
There are multiple possible physical explanations for this behavior, such as minimization of a Helfrich elastic energy of a receptor (i.e. a small patch or an oligomer) ~\cite{helfrich_elastic_1973}, induced by the underlying membrane,
\begin{equation}\label{eq:elastic-cell-membrane-energy}
E_{r}={\frac {k_{c}}{2}}(H-H_{0})^{2}+{\bar {k}}\,K.
\end{equation}
Here, $H_0$ is the preferred mean curvature of a receptor~\footnote{We assume our receptors to be a coarse-grained representation of small clusters of binding ligands, as there are orders of magnitude more receptors in real systems than in the ones simulated here}, and $H$ and $K$ are the mean and Gaussian curvature of the membrane, the latter of which can typically be neglected for cells~\cite{endres_polar_2009}.
Assuming $H_0 < H$ everywhere~\cite{koldso_lipid_2014}, energy is minimized when receptors localize at the highest negative mean curvature regions.
This approach shows agreement with experimental observations and explains why receptors are stable at a certain locations on the membrane~\cite{koldso_lipid_2014, draper_origins_2017,rosholm_membrane_2017,cai_t_2022}.
For a discussion of curvature-sensing proteins (as opposed to curvature-inducing proteins), see~\cite{noauthor_supplementary_nodate}.

\begin{figure}[!tb]
    \centering
    \subfloat[\label{fig:receptors_mean_curvature}]{\includegraphics[width=8cm]{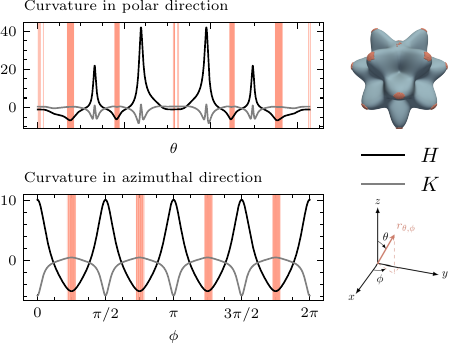}}
    \vspace{-1em}
    \subfloat[\label{fig:gradients_alignment}]{\includegraphics[width=8cm]{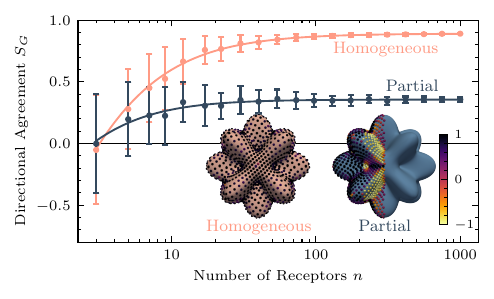}}
    \caption{
    Optimal receptor regions match those of minimal elastic energy on an spherical cell perturbed by the $Y_7^4$ harmonic.
    (a)~Mean curvature ($H$) in black and Gaussian curvature ($K$) in gray along the surface of the cell shown on the right, at fixed azimuthal angle $\phi=0\ \mathrm{rad}$ (top) and polar angle $\theta = 0.4\pi\ \mathrm{rad}$ (bottom).
    A single orange vertical line is placed at each angle where a receptor is found along the surface path.
    (b)~Direction alignment between the gradients from elastic energy and uncertainty minimization Eq.~\eqref{eq:alignment}, for homogeneously spread receptors (orange) and receptors located on half the surface (blue). Error bars show the standard deviation of 100 realizations, where receptor position is uniformly sampled at the entire or half cell surface.
    The diagrams show the individual alignment value (cosine similarity) of each receptor indicated by the color.
    }
    \label{fig:curvature_plot}
\end{figure}

Interestingly, in our study, deformation of the cell surface, which disrupts surface symmetry, results in redistribution of receptors typically to areas of low mean curvature, as shown in Fig.~\ref{fig:receptors_mean_curvature}.
In a curvature driven model, additional receptor-receptor interactions needs to be accounted for~\cite{endres_polar_2009}, whereas here clustering is not directly modeled; rather, it emerges through uncertainty minimization, with a cluster size set by the sensing correlation length.
Clusters do smoothly merge with each other as the effective receptor size $a$ increases~(see \cite{noauthor_supplementary_nodate}).
Yet, both models yield the same final localization for simple shapes.

Given that our resulting cluster locations match regions of minimal energy for simple shapes (Fig.~\ref{fig:receptors_mean_curvature}), we argue that gradient estimation is often optimized when receptors optimize for membrane curvature, driven by elastic membrane energy or alternative biophysical forces.
Thus, we find that this mechanistic paradigm (negative curvature maximization) that explains the physical reasons for cluster formation is intimately linked to a utilitarian evolutionary mechanism (enhanced gradient sensing), highlighting the \textsl{mechanical intelligence}~\cite{wan_origins_2021} of receptor localization.
Notably, although global information is necessary for optimizing gradient estimation, the curvature information is strictly local.
Thus, the agreement between receptor motion in the two mechanisms is only valid under certain conditions.
To quantify this, we calculate the alignment between receptor directions in both models using the cosine similarity between the gradients,
\begin{equation}\label{eq:alignment}
    S_G = \frac{\nabla \delta \ghati \cdot \nabla E_r}{\norm{\nabla \delta \ghati} \norm{\nabla E_r}}.
\end{equation}
As seen in Fig.~\ref{fig:gradients_alignment}, the alignment is strong when receptors are many and spread uniformly on the cell.
In contrast, when receptors are few and unevenly spread, e.g., only half the cell is covered in them, following the local energy gradients yield suboptimal gradient estimation, demonstrated by their misaligned gradient vectors.
Likewise, the correspondence between the two can fail locally e.g. for shapes that have small-scale localized areas of high (negative) curvature.
Nonetheless, their influence will be local, and as long as receptors start uniformly and spread, receptor relocalizations due to curvature will tend to improve spatial gradient sensing. 
We finally note that we consider a coarse-grained curvature, averaged over the scale of e.g. receptor oligomers of G-protein coupled receptors~\cite{gahbauer_membrate_2016}, and thus assume that curvature fluctuations below this scale will not affect receptor localization

\vspace{2em}
Moving beyond optimal \textsl{static} receptor locations, we study the consequences of membrane dynamics and the ensuing response of the receptor localization.
Many biological systems adapt to membrane deformations by relocating their sensors~\cite{koldso_lipid_2014, bernard_interplay_2024, potanin_kinetics_1994}, which can enhance chemotactic efficiency ~\cite{alonso_persistent_2024, wang_directional_2014}.
Thus, we allow receptors not only to move as a result of membrane deformations but also as an independent mechanism.
We model receptor movement as influenced by an effective potential that mirrors the gradient of the estimation variance, $U \sim \delta \hat{g}$, Eq.~\eqref{eq:ghat}.
This defines the best possible direction of motion, thus setting an upper bound on optimality compared to physical potentials such as those related to e.g. membrane curvature.
We assume that receptor motion in the viscous cell membrane is overdamped and restricted to a maximum speed $u$.
As the receptors are constrained to move on the two-dimensional star-convex cell surface, we formulate the gradient dynamics in spherical angles,
\begin{equation}\label{eq:motion-receptors}
    \partial_t \! \begin{pmatrix}\phi \\ \theta \end{pmatrix} = -\frac{u}{\max(\epsilon,  | \mathrm{d} \boldsymbol{r}|)} \nabla_{\phi, \theta} U,
\end{equation}
where the normalization from
\begin{equation}
    | \mathrm{d} \boldsymbol{r}|^2 = \sum_{q \in \{ x, y, z\}} \left( \frac{\partial q}{\partial \phi} \frac{\partial U}{\partial \phi} + \frac{\partial q}{\partial \theta} \frac{\partial U}{\partial \theta} \right)^2
\end{equation}
ensures that $u$ is the maximum speed in Cartesian coordinates, and $\epsilon$ is a small value that allows receptors to stop when the gradients are sufficiently small.

\begin{figure}[!htb]
    \centering
    \subfloat[\label{fig:evolution_catch}]{\includegraphics[width=8cm]{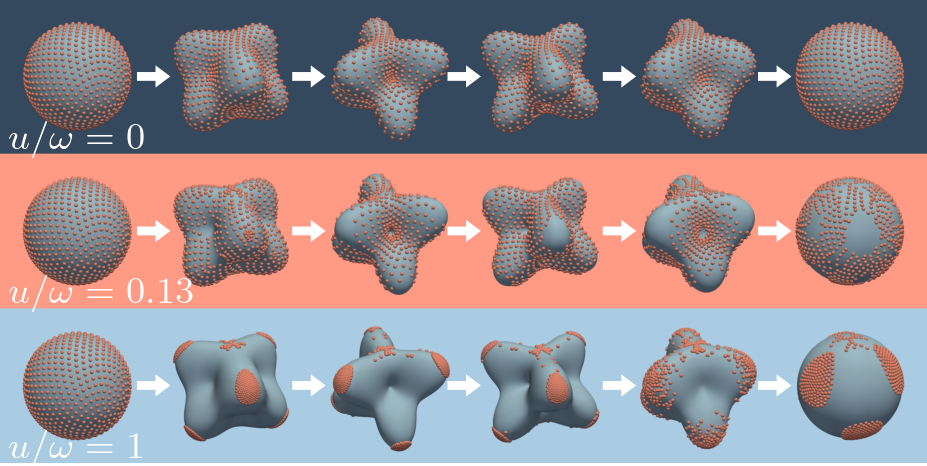}}
    
    \subfloat[\label{fig:speed_tradeoff}]{\includegraphics[width=8cm]{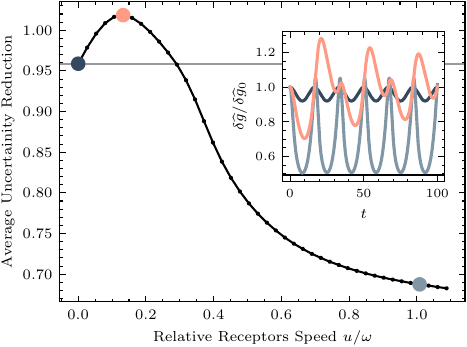}}
    
    \caption{Effects of reacting motile receptors to membrane dynamics.
    (a) Frames from three independent simulations of the motion form protrusions caused by the harmonic $Y_4^3$ perturbation, where the receptors are immobile (top), have low speed (middle) or move fast (bottom) as the membrane oscillates.
    (b) Average gradient estimation error on an oscillating motion where protrusions are grown and retracted (3 full oscillations).
    The inset shows the oscillations in error for the three cases displayed in (a) as the membrane changes shape, with the black line showing the reduction for \textsl{static} optimization.}
    \label{fig:moving_receptors}
\end{figure}

In order to study the effect of dynamical receptors, we perturb a perfect sphere to grow and retract protrusions, by defining the perturbation strength as an oscillating function $\alpha \propto \sin( \omega t)$, where $\omega$ is the speed of the membrane motion~(see~\cite{cavanagh_t_2022, tweedy_distinct_2013-1} for examples of oscillatory behavior in cells).
Thus, we perform simulations for increasing values of $u/\omega$, and observe the effects of finite-speed moving receptors on gradient estimation as the membrane of the cell surface is deformed~(Fig.~\ref{fig:moving_receptors}).
When receptors are immobile ($u/\omega=0$), and thus only move due to membrane changes, the performance of the estimation is only slightly affected by the perturbations as seen in Fig.~\ref{fig:speed_tradeoff}.
Paradoxically, when receptors are allowed to move at low speeds ($u/\omega=0.13$), the average accuracy decreases compared to when the receptors are not motile.
This is because the receptors move according to instantaneous membrane shape information, and are not fast enough to reach the tip of protrusions and form clusters before those tips have disappeared again.
As the speed of the receptors becomes faster, clusters form, and the performance converges towards the same optimal spatial gradient sensing as that of static shapes.
We note that Eq. \eqref{eq:motion-receptors} is the simplest choice for how receptors will follow cell shape changes, with a more precise formulation depending on a microscopic description of the membrane.
For all models, receptor clustering can lag behind their optimal positions, but the quantitative values depend on the specifics of the microscopic model.

\vspace{2em}
The morphology of the observed clusters has until now been a conglomerate of receptors spread isotropically.
This is due to the fact that the environment has been symmetric, i.e. we have assumed the cell to be static and in a shallow gradient.
However, chemotactic cells are naturally motile --- motility being the typical reason to optimize gradient estimation.
To that end, we study how the resulting cluster formation is affected when flow is accounted for; this flow being the result either of external fluid motion or cell motility.

With flow, chemical molecules not only diffuse but are also advected along the flow.
This changes the covariance between receptors.
Under certain simplifying assumptions (see~\cite{noauthor_supplementary_nodate} for derivation), we find
\begin{equation}\label{eq:covariance-flow}
\hat \Sigma_{i,j} = \frac{2 a^3 V}{D\tau(5 a + 6 \Delta)}e^{\frac{-\norm{\boldsymbol{v}}\sqrt{a^2+\Delta^2}}{2D}}(\x_i e^{-\lambda} + \x_j e^{\lambda})  \cdot \g, 
\end{equation}
where $\lambda=\boldsymbol{v}\cdot \boldsymbol{\Delta}/2D$, and $\boldsymbol{v}$ is the flow, which we take to be the Stokes flow that results from moving the deformed cells at a constant speed.
We compute this numerically using the boundary element regularized Stokeslet method~\cite{cortez_method_2001, smith_boundary_2009}.

\begin{figure}[htb]
    \centering
    \includegraphics[height=8cm]{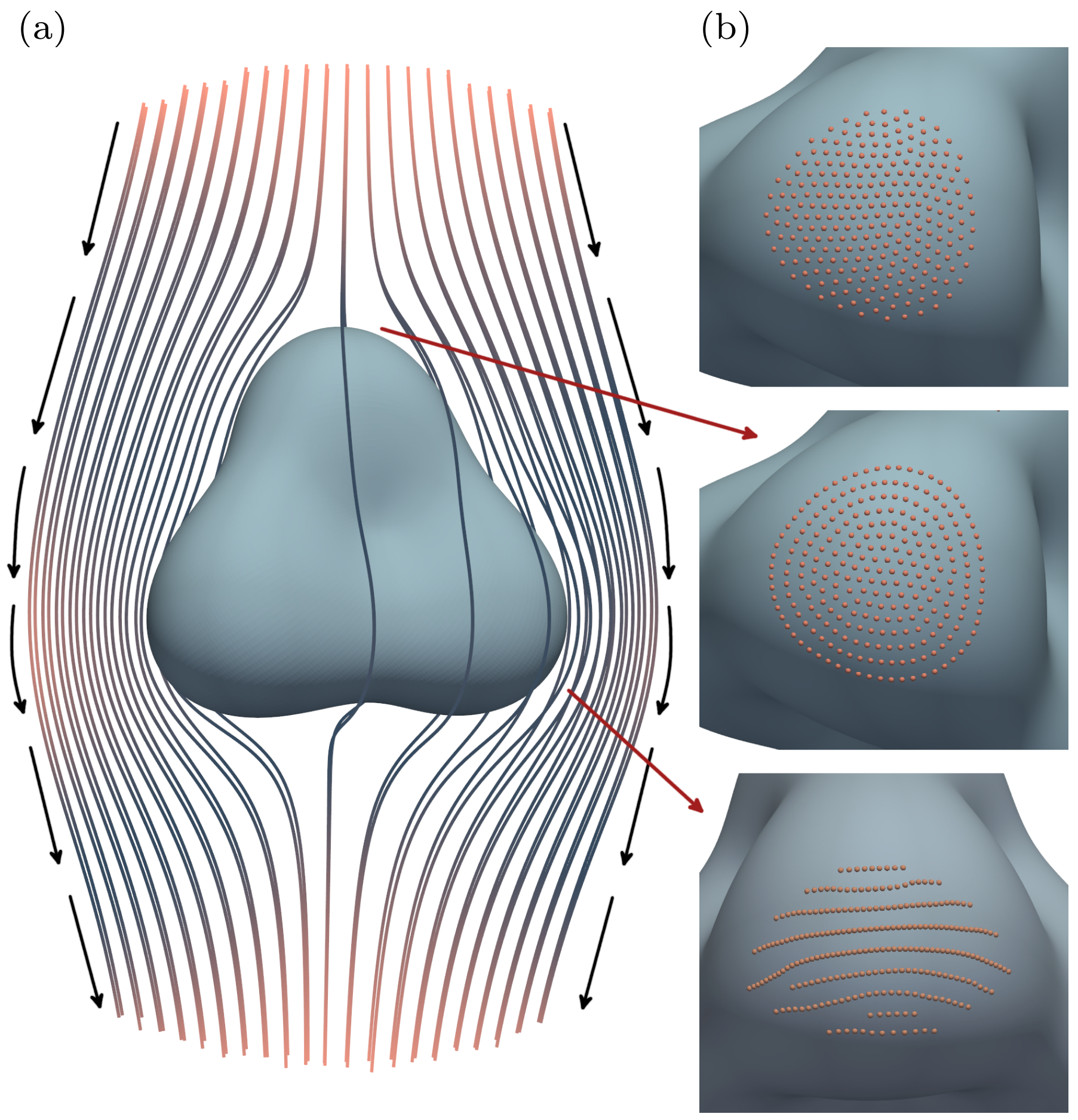}
    \caption{
    Cluster formation in the presence of flow for a cell perturbed by the $Y_3^2$ harmonic.
    (a)~Streamlines of the simulated Stokes flow around the cell.
    (b)~Resulting cluster on an environment without flow (top), cluster formation under the effect of front-facing fluid flow (middle) and cluster morphology for a side protrusion affected by flow (bottom).
    }
    \label{fig:flow-clusters}
\end{figure}

We observe that the cluster morphology is drastically affected by the inclusion of flow~(Fig.~\ref{fig:flow-clusters}).
Interestingly, because of the higher correlation of the receptors in the direction of the fluid flow, given by Eq.~\ref{eq:covariance-flow}, receptors in clusters are no longer isotropically spread.
Instead, they converge to form lines perpendicular to the flow field.

\vspace{1em}
In this Letter, we found the optimal placement of cell-surface receptors by developing a theoretical model that minimizes uncertainty in gradient estimation.
Our results show that, in spherical cells, optimal receptor location is uniform across the surface, but once symmetry is broken through shape perturbations, clusters of receptors naturally emerge in regions of high negative mean curvature.
These clusters form without explicitly including curvature information, yet are, under certain conditions, consistent with the results of (negative) curvature maximization.
This supports the idea that, even though the receptor distribution is governed by biophysical principles, the final distribution agrees with an evolutionary trait to maximize the sensing accuracy.

We observed that when accounting for receptor localization, the effect of shape deformations on gradient estimation accuracy is massively improved,
with the effect of receptor localization being more than an order of magnitude larger than the effect stemming from the shape deformation on its own.
When shape deformations are dynamic, we showed that surface receptor motility can maintain the massive improvement in estimation accuracy, provided that receptors move \textsl{sufficiently} fast.
This would indeed be the case, for example, during cell migration of social amoeba~\cite{tweedy_distinct_2013} and T cells~\cite{cavanagh_t_2022} with observed shape deformations on time scales of $3-4\ \text{min}$ and receptor diffusivities of $\sim0.1\upmu \text{m}^2/\text{s}$~\cite{nguyen_how_2015}, as long as there are sufficient receptors to avoid long distances for diffusion~($\geq 1\upmu \text{m}$).
In contrast, if the receptors were too slow, receptor motility can have adverse effects on accuracy of estimation.
Note that gradient sensing, which occurs over a time-scale of seconds~\cite{endres_accuracy_2008}, is much faster than the time required by the receptors to relocalize. 
Finally, we have shown how optimal clustering is affected by fluid flow, leading to separation of receptors in the direction of the flow to reduce receptor-receptor correlations.
Taken together, our work shows that cells can significantly improve the accuracy of spatial sensing by actively regulating receptor placement and clustering.

\begin{acknowledgments}
This work has received funding from the Novo Nordisk Foundation, Grant Agreement No. NNF20OC0062047.
\end{acknowledgments}

\bibliography{references}

\end{document}